\documentclass[aps,prd,reprint,twocolumn,preprintnumbers,superscriptaddress,nofootinbib]{revtex4-1}

\usepackage{amsmath}
\usepackage{graphicx}
\usepackage{float}
\usepackage[colorlinks=True, citecolor=blue, urlcolor=blue, linkcolor=blue]{hyperref}

\usepackage{amssymb}
\usepackage[normalem]{ulem}

\begin{document}

\title{Point cloud-based diffusion models for the Electron-Ion Collider}

\author{Jack Y. Araz}
\email{jack.araz@stonybrook.edu}
\affiliation{Center for Nuclear Theory, Department of Physics and Astronomy,
Stony Brook University, New York 11794, USA}
\affiliation{Thomas Jefferson National Accelerator Facility, Newport News, VA 23606, USA}
\affiliation{Department of Physics, Old Dominion University, Norfolk, VA 23529, USA}

\author{Vinicius Mikuni}
\email{vmikuni@lbl.gov}
\affiliation{National Energy Research Scientific Computing Center, Berkeley Lab, Berkeley, CA 94720, USA}

\author{Felix Ringer}
\email{felix.ringer@stonybrook.edu}
\affiliation{Center for Nuclear Theory, Department of Physics and Astronomy,
Stony Brook University, New York 11794, USA}
\affiliation{Thomas Jefferson National Accelerator Facility, Newport News, VA 23606, USA}
\affiliation{Department of Physics, Old Dominion University, Norfolk, VA 23529, USA}

\author{Nobuo Sato}
\email{nsato@jlab.org}
\affiliation{Thomas Jefferson National Accelerator Facility, Newport News, VA 23606, USA}

\author{Fernando Torales Acosta}
\email{ftoralesacosta@lbl.gov}
\affiliation{Physics Division, Lawrence Berkeley National Laboratory, Berkeley, CA 94720, USA}

\author{Richard Whitehill}
\email{rwhit058@odu.edu}
\affiliation{Department of Physics, Old Dominion University, Norfolk, VA 23529, USA}

\preprint{JLAB-THY-24-4224s}


\begin{abstract}
At high-energy collider experiments, generative models can be used for a wide range of tasks, including fast detector simulations, unfolding, searches of physics beyond the Standard Model, and inference tasks. In particular, it has been demonstrated that score-based diffusion models can generate high-fidelity and accurate samples of jets or collider events. This work expands on previous generative models in three distinct ways. First, our model is trained to generate entire collider events, including all particle species with complete kinematic information. We quantify how well the model learns event-wide constraints such as the conservation of momentum and discrete quantum numbers. We focus on the events at the future Electron-Ion Collider, but we expect that our results can be extended to proton-proton and heavy-ion collisions. Second, previous generative models often relied on image-based techniques. The sparsity of the data can negatively affect the fidelity and sampling time of the model. We address these issues using point clouds and a novel architecture combining edge creation with transformer modules called Point Edge Transformers. Third, we adapt the foundation model OmniLearn, to generate full collider events. This approach may indicate a transition toward adapting and fine-tuning foundation models for downstream tasks instead of training new models from scratch.
\end{abstract}


\maketitle

\section{Introduction}

High-energy collider experiments offer unique opportunities to probe the internal dynamics of protons and nuclei, study emergent phenomena such as hadronization, and search for physics beyond the Standard Model of particle physics. By analyzing the particles observed in detectors centered around the scattering vertex, it is possible to infer the dynamics of particles at subatomic scales. The next-generation experiment will be the future Electron-Ion Collider (EIC)~\cite{AbdulKhalek:2021gbh}, where high-luminosity electron-proton/nucleus scattering will be studied at center-of-mass (CM) energies up to $\sqrt{s}=140$~GeV. Analyzing vast amounts of recorded collider data is a challenging task where machine learning tools are expected to have a significant impact on the experimental and theoretical workflows. Example applications include detector design, inference tasks, searches of BSM physics, fast detector simulations, jet classification, and unfolding. Similar considerations apply to proton-proton and heavy-ion collisions at RHIC and the LHC. For recent results, see Refs.~\cite{2013arXiv1312.6114K,Goodfellow:2014,DBLP:journals/corr/Sohl-DicksteinW15,DBLP:journals/corr/abs-2006-11239,Kasieczka:2017nvn,Cai:2021hnn,Datta:2017rhs,Komiske:2018cqr,Heimel:2018mkt,Dreyer:2021hhr,Bellagente:2019uyp,Andreassen:2019cjw,Alanazi:2020jod,Alghamdi:2023emm,Huang:2023kgs,Lee:2022kdn,Butter:2019cae,Gao:2020zvv,Danziger:2021eeg,Butter:2022rso,Cirigliano:2021img,Nachman:2020lpy,Atkinson:2022uzb,Alghamdi:2023emm,Scheinker:2024anx,Andreassen:2020nkr,Finke:2021sdf,Fraser:2021lxm,Araz:2022zxk,Sengupta:2023vtm,Morandini:2023pwj,Birk:2024knn} and references therein.

\begin{figure*}[t]
\label{fig:architecture}
    \centering
    \includegraphics[width = 0.95\textwidth]{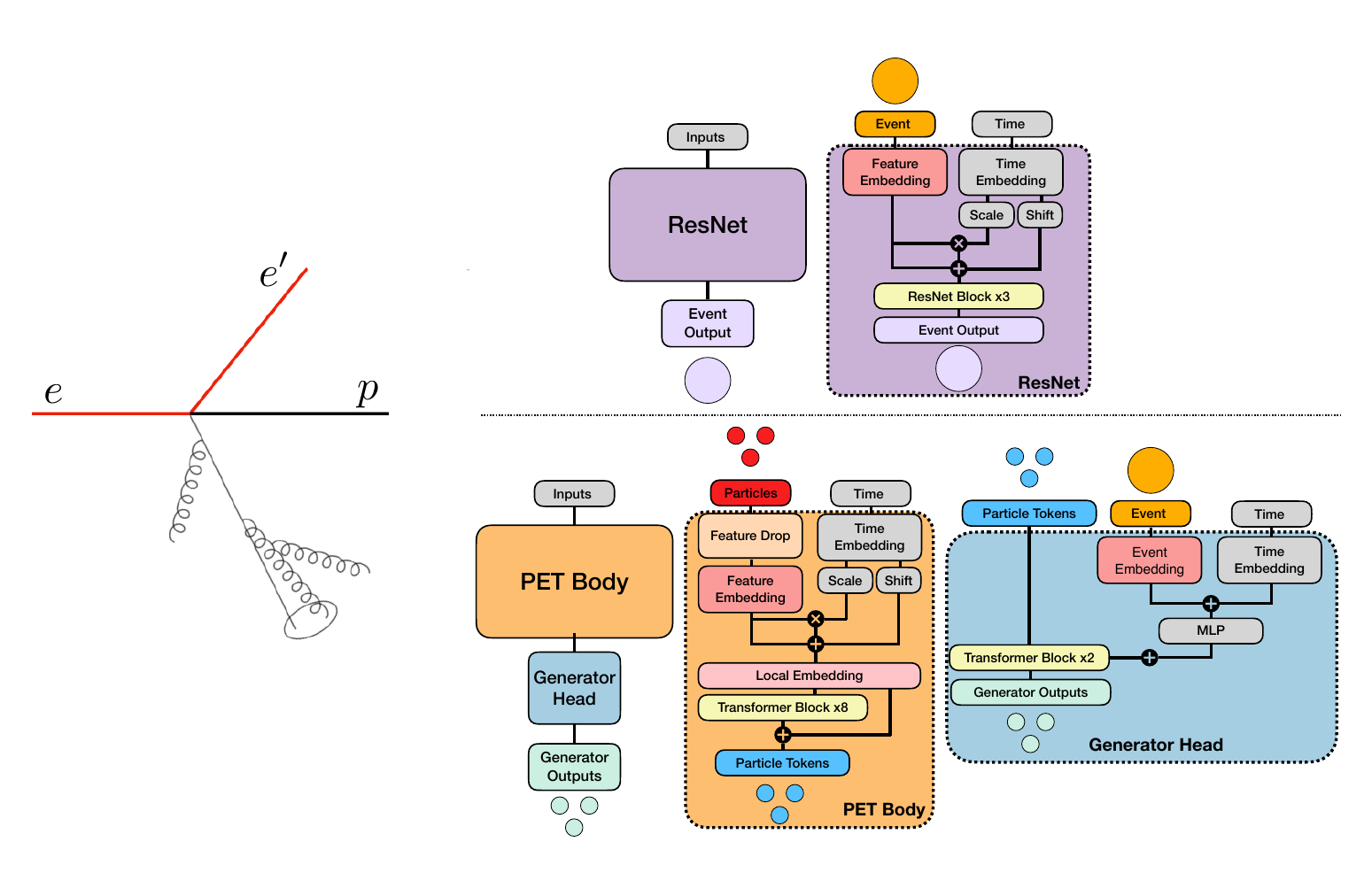}
\caption{Left: Electron-proton scattering event $e+p\to e'+X$. Right: Model architecture adapted from the foundation model OmniLearn~\cite{Mikuni:2024qsr}. The final model is composed of two diffusion models: One that generates the scattered electron and the event properties, such as the multiplicity (top), and a second model that generates all other particles in the event with their kinematics (bottom).~\label{fig:architecture}}
\end{figure*}

Some of the key tools to advance different areas of collider phenomenology are generative models that can be trained to generate full collider events. Various architectures have been trained in the past to generate collider events or jets including GANs~\cite{deOliveira:2017pjk,Paganini:2017dwg,Alanazi:2020klf,Kansal:2021cqp,Buhmann:2023pmh}, variational autoencoders~\cite{Touranakou:2022qrp},  normalizing flows~\cite{Kach:2022qnf,Verheyen:2022tov} and score-based diffusion models~\cite{DBLP:journals/corr/abs-1907-05600,Mikuni:2022xry,Mikuni:2023dvk,Leigh:2023toe,Butter:2023fov,Acosta:2023zik,Leigh:2023zle,Buhmann:2023pmh,Amram:2023onf,Buhmann:2023zgc,Imani:2023blb}. In particular, diffusion models have been demonstrated to produce high-fidelity samples. While the sampling time is typically relatively slow compared to GANs, it has been improved significantly using techniques such as distillation~\cite{Mikuni:2023tqg,Mikuni:2023dvk}. Score-based diffusion models learn an approximation of the score function or the gradients of the logarithm of the data probability. This approximation is then used during sampling to transform a simple distribution, such as a Gaussian distribution, into complex collider data. In Ref.~\cite{Devlin:2023jzp}, a first diffusion-based model was developed for EIC events based on pixelated images. Since $>99\%$ of the pixels are empty and the distributions of different observables can fall steeply toward their kinematic endpoints, a suitable remapping of the input variables was required. Due to its relevance in Deep Inelastic Scattering (DIS), the modeling of the scattered electron kinematics plays a critical role in electron-proton collisions that requires special attention. See also Ref.~\cite{Go:2024xor} where an image-based diffusion model was developed for heavy-ion collisions.

In this work, we extend previous results by developing a point cloud-based diffusion model for EIC events. By using point clouds and a novel architecture that combines edge creation with transformer modules, termed Point Edge Transformers (PET), we achieve significant improvements compared to the diffusion model of Ref.~\cite{Devlin:2023jzp}. In particular, we focus on success metrics such as the shape of different kinematic distributions as well as the event-wide conservation of momentum and discrete quantum numbers. We adapt the pre-existing foundation model OmniLearn~\cite{Mikuni:2024qsr} to generate full EIC events. OmniLearn was initially developed for both classification and generation tasks in the context of jet physics at the LHC. To generate EIC events, including full Particle IDentification (PID), we use a two-step generation process. As a first step, the scattered electron kinematics are generated. Second, the remaining particles in the event are conditioned on the electron kinematics. We expect similar multi-step generative processes may also improve the generation of full events in different collision systems. While we train the model developed here from scratch instead of fine-tuning the foundation model, our approach is closely related to OmniLearn. Our results may, therefore, point toward a transition toward adapting foundation models for different downstream tasks at collider experiments.

The remainder of this paper is organized as follows. In section~\ref{sec:model}, we describe the score-based diffusion model for EIC events developed in this work employing a point cloud data representation and the PET architecture. In section~\ref{sec:num}, we consider several metrics to evaluate the performance of the diffusion model. We consider different particle distributions and observables as well as event-wide constraints such as momentum and baryon number conservation. We conclude and present an outlook in section~\ref{sec:conclusions}.

\section{Point cloud-based diffusion models~\label{sec:model}}

\begin{figure*}[t]
    \centering
    \includegraphics[width=1\textwidth]{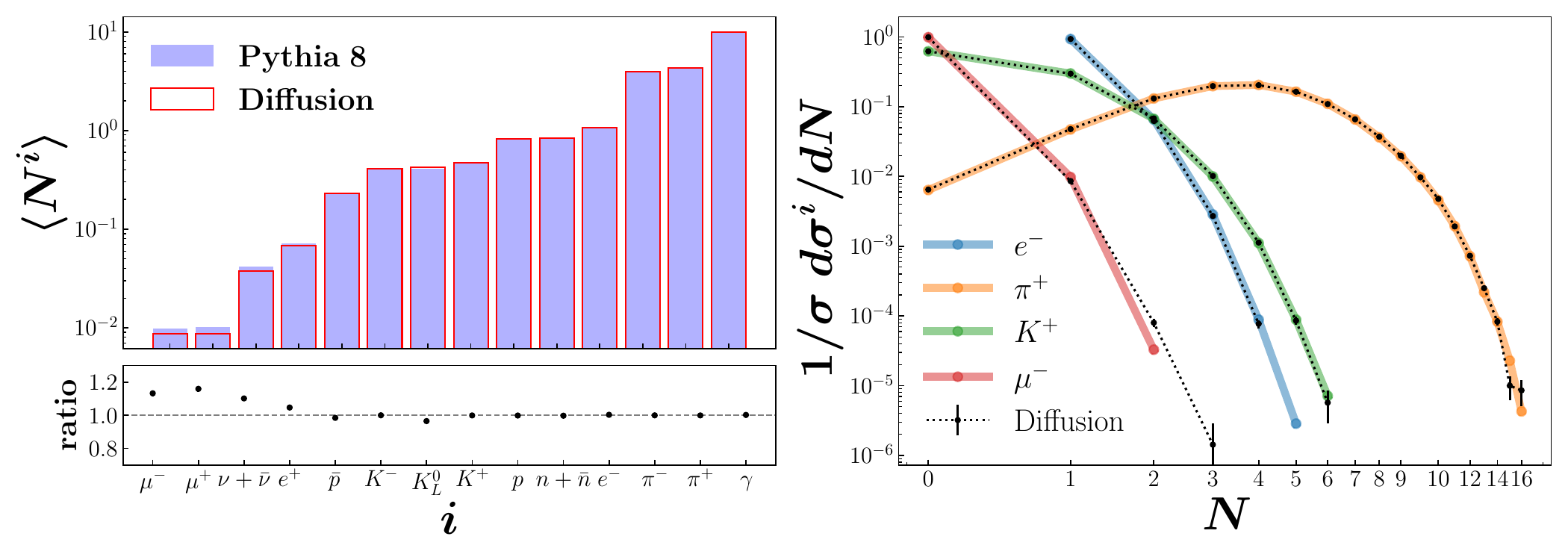}
    \caption{Left: Average particle multiplicities produced by the diffusion model compared to the \textsc{Pythia8} training data. Right: Comparison of the event-wide particle multiplicity distributions for electrons $e^-$, pions $\pi^+$, kaons $K^+$, and muons $\mu^-$.}
    \label{fig:AverageMultiplicities}
\end{figure*}

We start by reviewing the generation of EIC events used to train the diffusion model. We then describe the model architecture and the two-step diffusion process used to generate electron-proton events.

\subsection{Event generation and data representation}

We generate electron-proton scattering events using \textsc{Pythia8}~\cite{Sjostrand:2014zea} at a representative CM energy for electron-nucleus collisions at the EIC $\sqrt{s}=105$~GeV. We avoid the low-$Q^2$ photoproduction region by imposing a cut of $Q^2>25$~GeV$^2$. We include the following list of stable particles in the data set 
\begin{equation}
\label{eq:particles-included}
    e^\pm,\mu^\pm,\nu+\bar\nu,\pi^\pm,\pi^0,K^\pm,K_L^0,p,\bar p, n+\bar n,\gamma\,.
\end{equation}
We include all particles in the rapidity range $|y|<5$, and we do not impose a lower cut on the transverse momentum. Note that here, $\nu+\bar\nu$ and $n+\bar n$ are combined due to experimental limitations in distinguishing them.

For each particle $i$ in the event, we record its transverse momentum $p_{Ti}$, rapidity $y_i$, azimuthal angle $\phi_i$, and PID$_i$. In addition, we consider the dimensionless quantity
\begin{equation}
    \tilde{z}_i=\frac{2M_{Ti}}{\sqrt{s}}\cosh y_i \,.
\label{eq:z}
\end{equation}
Here, $M_{Ti}^2=p_{Ti}^2+m_i^2$ is the transverse mass, and $m_i$ is the mass of the particle. This variable is of particular interest as it satisfies
\begin{equation}\label{eq:mtmconservation}
    \sum_{i\in{\rm event}}\tilde{z}_i=2\,,
\end{equation}
in the CM frame due to event-wide momentum conservation.
In the limit of massless particles $\tilde{z}_i$ reduces to
\begin{equation}
    z_i = \frac{2 p_{Ti}}{\sqrt{s}}\cosh{\eta_i}\,,
\end{equation}
where $\eta_i=-\ln{\tan{\theta_i/2}}$ is the $i^{\rm th}$ particle's pseudorapidity.
While the relation between a massive particle's rapidity and pseudorapidity is somewhat intricate, one can convert between the variables $\tilde{z}_i$ and $z_i$ using the relation $\tilde{z}_i=\sqrt{z_i^2+4m_i^2/s}$.

\subsection{Model architecture: Point Edge Transformer}

This work extends the generalized machine learning model, OmniLearn \cite{Mikuni:2024qsr}, designed for analyzing data from particle physics experiments. The model processes inputs consisting of particles and event-level information such as the particle multiplicity and is conditioned on a diffusion time parameter $t$ that determines the perturbation level applied to the data. In particular, for time $t$ we apply a perturbation to data $x$ such that $x(t) = \alpha(t)x + \sigma(t)\epsilon $, with $\epsilon\sim \mathcal{N}(0,1)$ and perturbation parameter $\alpha(t) = \cos(\pi t/2)$ and $\sigma(t) = \sin(\pi t/2)$. The role of the network is then to predict a velocity parameter $v(t) = \alpha(t)\epsilon - \sigma(t)x$ by receiving as inputs the perturbed data, the time value, and any additional event-level information available. The time information for the diffusion process, as done in previous diffusion models for collider physics~\cite{Mikuni:2022xry,Mikuni:2023dvk,Mikuni:2023tqg}, is encoded to a higher dimensional space using a time embedding layer. This embedding layer utilizes Fourier features~\cite{tancik2020fourier} and is further processed by two multi-layer perceptrons (MLPs) employing a GELU activation function~\cite{hendrycks2016gaussian}.

\begin{figure*}
    \centering
    \includegraphics[width=\linewidth]{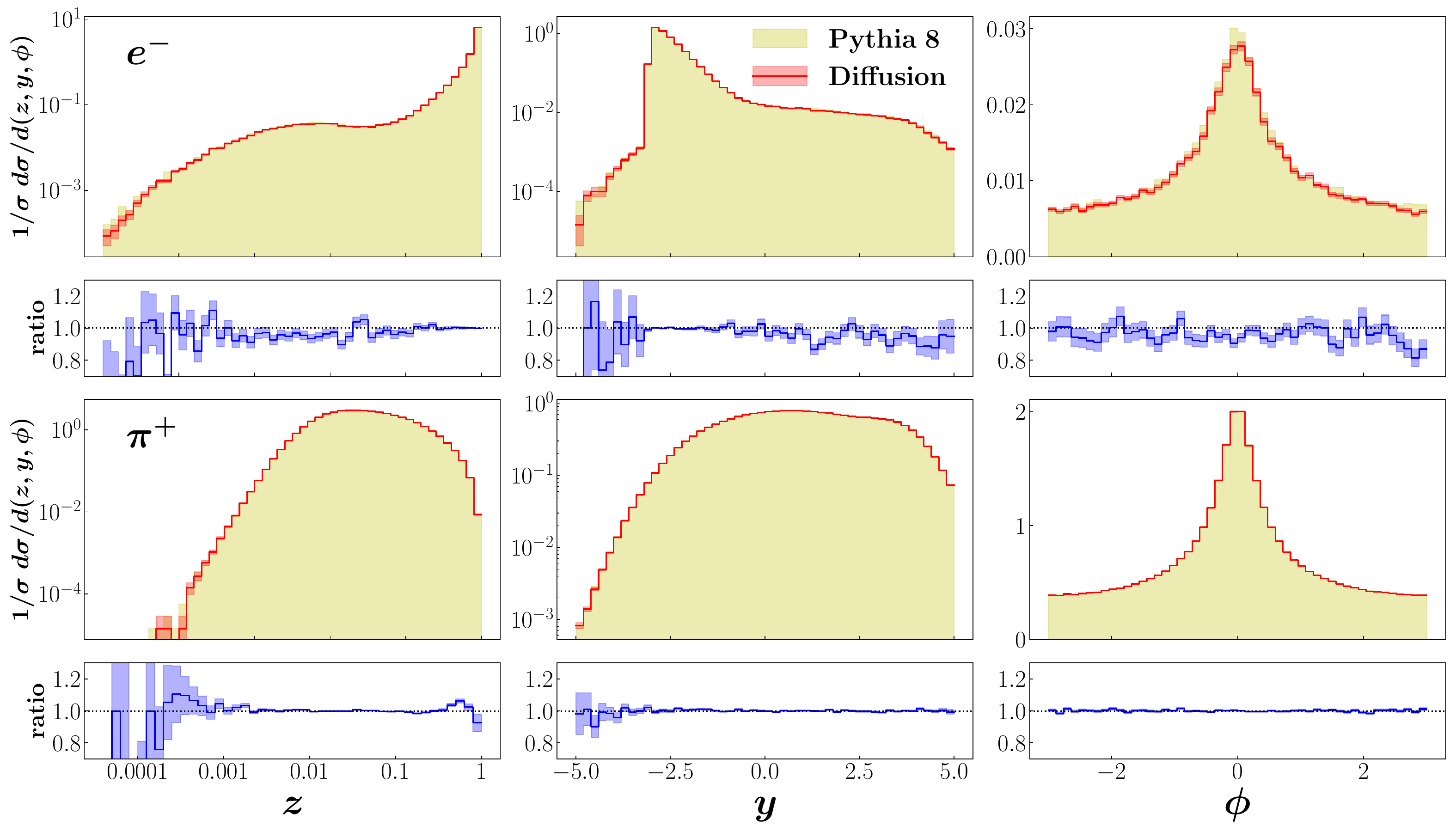}
    \caption{Left to right: Kinematic distributions for the rescaled momentum variable $z$, rapidity $y$, and azimuthal angle $\phi$ (relative to the scattered leading electron). We show the diffusion model results along with the \textsc{Pythia8} training data for electrons (top row) and pions (bottom row). The shaded red uncertainties show the statistical errors of the diffusion model. The blue error bands in the ratio plots include the statistical uncertainties from the diffusion model and \textsc{Pythia8}.}
    \label{fig:Kinematics}
\end{figure*}

The generative model designed to produce the scattered electron kinematic information is based on a fully-connected architecture incorporating multiple skip connections. Specifically, the model employs three \textsc{ResNet}~\cite{he2016deep} blocks, where each residual layer is connected to the output of a two-layer network through a skip connection.  The model is designed to generate particles and then integrates the time-related data with particle-specific information, which includes both the kinematics of each particle and their PID after the perturbation. These inputs are transformed into a higher dimensional space using a feature embedding composed of two MLP layers. 
Prior to the transformer block -- which is responsible for processing data in a manner that considers the relationships between particles -- we insert a positional token. This token encodes the geometric context surrounding each particle in the event, aiding the transformer in understanding local particle arrangements. Although transformers are capable of capturing broad correlations among particles, adding local geometric data typically enhances the model’s performance by creating a latent representation aware of particle distances~\cite{Mikuni:2021pou}.
The local encoding is constructed using dynamic graph convolutional network (DGCNN)~\cite{DBLP:journals/corr/abs-1801-07829} layers, which define each particle’s neighborhood through a k-nearest neighbor algorithm, set to include ten neighbors in this work. The distances between these neighbors are measured in the specific rapidity-azimuthal angle space. For each of the k-neighbors, edge features are defined by concatenating the particle features with the subtraction between those features and the features of each respective neighbor. These edge features are then processed by a multi-layer perception (MLP), followed by an average pooling operation performed across the dimensions of the neighbors.

\subsection{Two-step diffusion and training}

We adopt the two-model strategy implemented in Ref.~\cite{Mikuni:2023dvk}. See Fig.~\ref{fig:architecture} for an illustration of the model architecture developed here. The first model is trained to exclusively learn event-level features that are then utilized as conditional information for a second diffusion model that processes particles as inputs. Most important for this process is the total number of particles in the event, $N$, which is learned by the first diffusion model. $N$ is then shared with the second diffusion model that then generates $N$ particles and all their features for that event. Up to 50 particles are saved per event to be used during training, the maximum of all \textsc{Pythia8} events in the training sample. 

In addition to the global event variables such as multiplicity, the first model is also tasked to learn the kinematic distribution of the scattered electron in the event. This information is then used to generate the particle candidates: instead of generating the full four-momentum of each particle $i$ we use the electron $e^-$ to generate particles in relative coordinates, learning instead $\phi_i-\phi_e$, $y_i + y_e$, and $p_{Ti}/p_{Te}$. This choice of coordinates is invariant under rotations in the $y-\phi$ plane and improves the model generalization. The set of particle features learned by the second diffusion model is:
\begin{equation}
    \log_{10}(p_{Ti}/p_{Te}),\ y_i+y_e,\ \phi_i-\phi_e,\ \log_{10}(\tilde z_i),\ C_i,\ \mathrm{PID}_i,
\label{eq:particle_features}
\end{equation}
where $C_i$ is the particle charge, and $\tilde z_i$ is given in Eq.~(\ref{eq:z}). The ranges of $p_{Ti}/p_{Te}$ and $\tilde z_i$ still span several orders of magnitude, so the logarithm is taken for better normalization.

By conditioning the full event distributions on the dominant flow of momentum, we expect that the multi-step approach developed here can be extended to other collision systems such as $eA$, $e^+e^-$, $pp$, and heavy-ion collisions. We leave a more detailed exploration for future work.

The training is carried out on the Perlmutter Supercomputer~\cite{Perlmutter} using 128 GPUs simultaneously with the Horovod~\cite{sergeev2018horovod} package for data distributed training. A local batch of size 256 is used with model training up to 200 epochs. OmniLearn is implemented in \textsc{TensorFlow}~\cite{tensorflow} with \textsc{Keras}~\cite{keras} backend. The cosine learning rate schedule~\cite{DBLP:journals/corr/LoshchilovH16a} is used with an initial learning rate of $3\times10^{-5}$, increasing to $3\sqrt{128}\times10^{-5}$ after three epochs and decreasing to $10^{-6}$ until the end of the training. The \textsc{Lion} optimizer~\cite{chen2024symbolic} is used with parameters $\beta_1 = 0.95$ and $\beta_2 = 0.99$. The \textsc{PET} body model has 1.3M trainable weights, while the generator head has 416k trainable parameters. 

\section{Numerical results~\label{sec:num}}

In this section, we consider different kinematic distributions as benchmarks to assess the performance of the point cloud-based diffusion model. In addition, we consider several event-wide constraints and we quantitatively assess the improvement compared to the image-based diffusion model for electron-proton scattering events presented in Ref.~\cite{Devlin:2023jzp}.

\subsection{Kinematic distributions with full PID}

We start by analyzing the average particle multiplicities per event. In the left panel of Fig.~\ref{fig:AverageMultiplicities}, we show the results from the diffusion model compared to \textsc{Pythia8} for all particle species. Overall, the diffusion model performs better for particles with higher average multiplicity. For several of the most frequently produced particles, the average yield from the diffusion model agrees with \textsc{Pythia8} within the statistical uncertainties. For muons, which have the lowest average yield of the particles considered here, we observe differences of $\lesssim 20\%$. This can be attributed to the fact that the muon yield is three orders of magnitude lower than, for example, the photon multiplicity. Instead of considering only the average multiplicities, we plot the particle multiplicity distributions for four representative examples in the right panel of Fig.~\ref{fig:AverageMultiplicities}. Overall, we observe good agreement between the diffusion model and the \textsc{Pythia8} results. The distributions fall over multiple orders of magnitude for large multiplicities and we observe small differences only in the tails of the distributions.

\begin{figure}
    \centering
    \includegraphics[width=\linewidth]{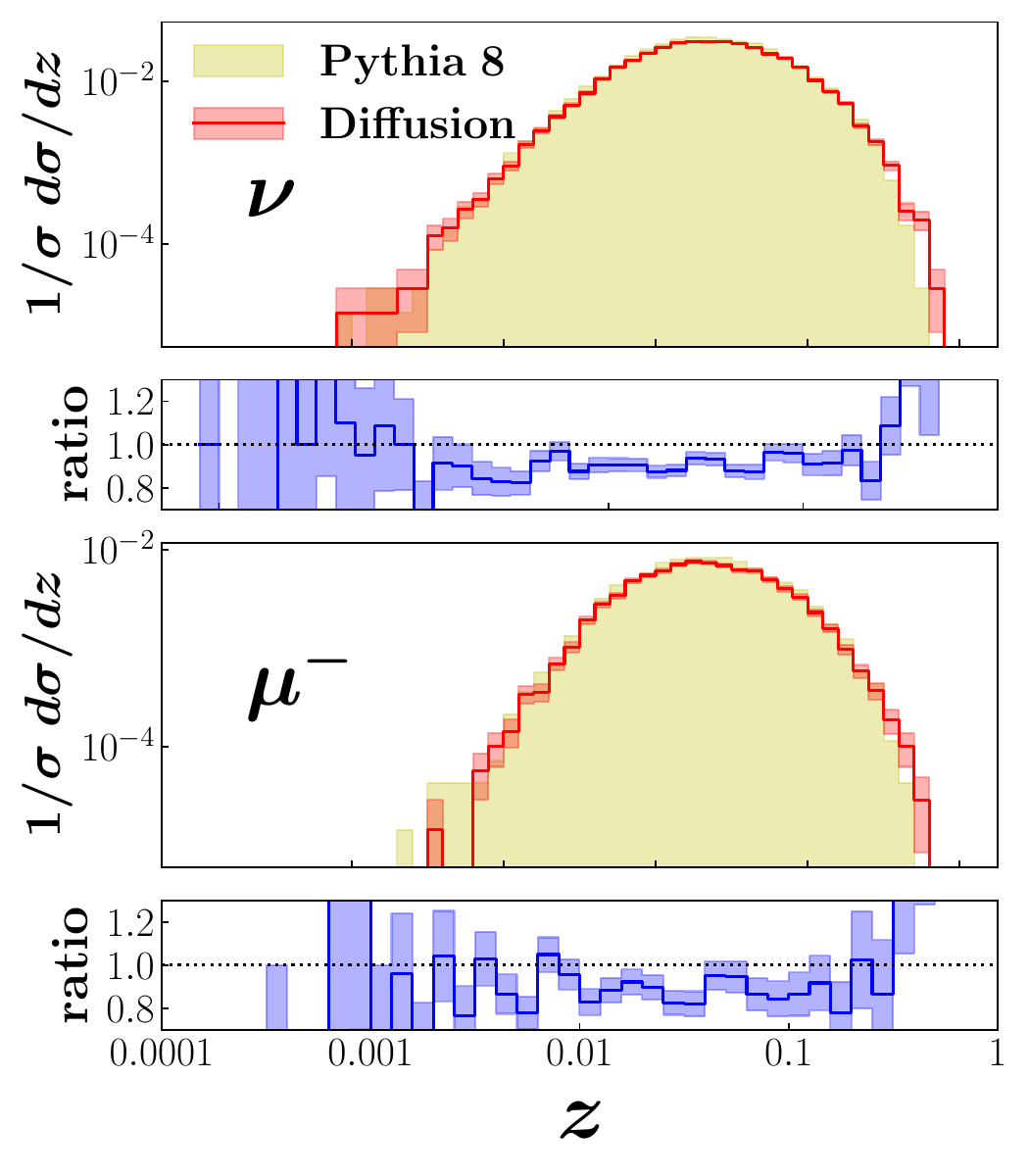}
    \caption{Distributions of the rescaled momentum variable $z$ for neutrinos (top) and muons (bottom) from the diffusion model and \textsc{Pythia8}.}
    \label{fig:Kinematics2}
\end{figure}

As a next step, we consider the distribution of different kinematic variables. Since the distributions exhibit rather different features, we choose electrons $e^-$ and pions $\pi^+$ as representative examples. The results from the diffusion model compared to \textsc{Pythia8} are shown in Fig.~\ref{fig:Kinematics}. We show histograms for the rescaled momentum fraction $z$, the rapidity $y$, and the azimuthal angle $\phi$. For both particle species, the azimuthal angle is considered relative to the scattered leading electron. We observe good agreement in the bulk of the distribution and smaller deviations toward the endpoints where low statistics lead to larger uncertainties that are, however, statistically distributed around the target result. We note that these results constitute a significant improvement compared to the diffusion model for electron-proton events presented in Ref.~\cite{Devlin:2023jzp}. To further evaluate the performance of the diffusion model, we consider the kinematic distributions of muons and neutrinos, which are the particles with the lowest average event multiplicities, see Fig.~\ref{fig:AverageMultiplicities}. As an example, we show a comparison of the $z$-distributions in Fig.~\ref{fig:Kinematics2}. As expected, while the agreement between the diffusion model and \textsc{Pythia8} is slightly worse compared to the distributions for electrons and pions in Fig.~\ref{fig:Kinematics}, we find overall satisfactory results.

Next, we consider kinematic variables that are particularly relevant for the analysis of electron-proton scattering data. First, we consider the Deep Inelastic Scattering (DIS) cross-section, which is differential in the scaling variable Bjorken $x$ and the photon virtuality
\begin{equation}\label{eq:DISVar}
x  = \frac{Q^2}{2 P \cdot q} \,, \quad Q^2  = -q^2=-\left(k-k^{\prime}\right)^2\,.
\end{equation}
Here $k,k'$ are the four momenta of the incoming and and outgoing electron, respectively, and $P$ denotes the incoming proton momentum. Second, we consider Semi-Inclusive DIS (SIDIS), where the following two additional variables are typically defined
\begin{equation}\label{eq:SIDISVar}
    z_{h} = \frac{P \cdot P_h}{P \cdot q}\,,\quad q_T=\frac{p_{Th}}{z_{h}}\,.
\end{equation}
Here, $P_h$ is the four-momentum of an observed final-state hadron, and $p_{Th}$ is its transverse momentum in the Breit frame. In the target rest frame,  $z_{h}$ is the hadron energy over the photon virtuality. See Ref.~\cite{Bacchetta:2006tn} for frame-independent definitions of the relevant variables listed above.

\begin{figure}
    \centering
    \includegraphics[width=\linewidth]{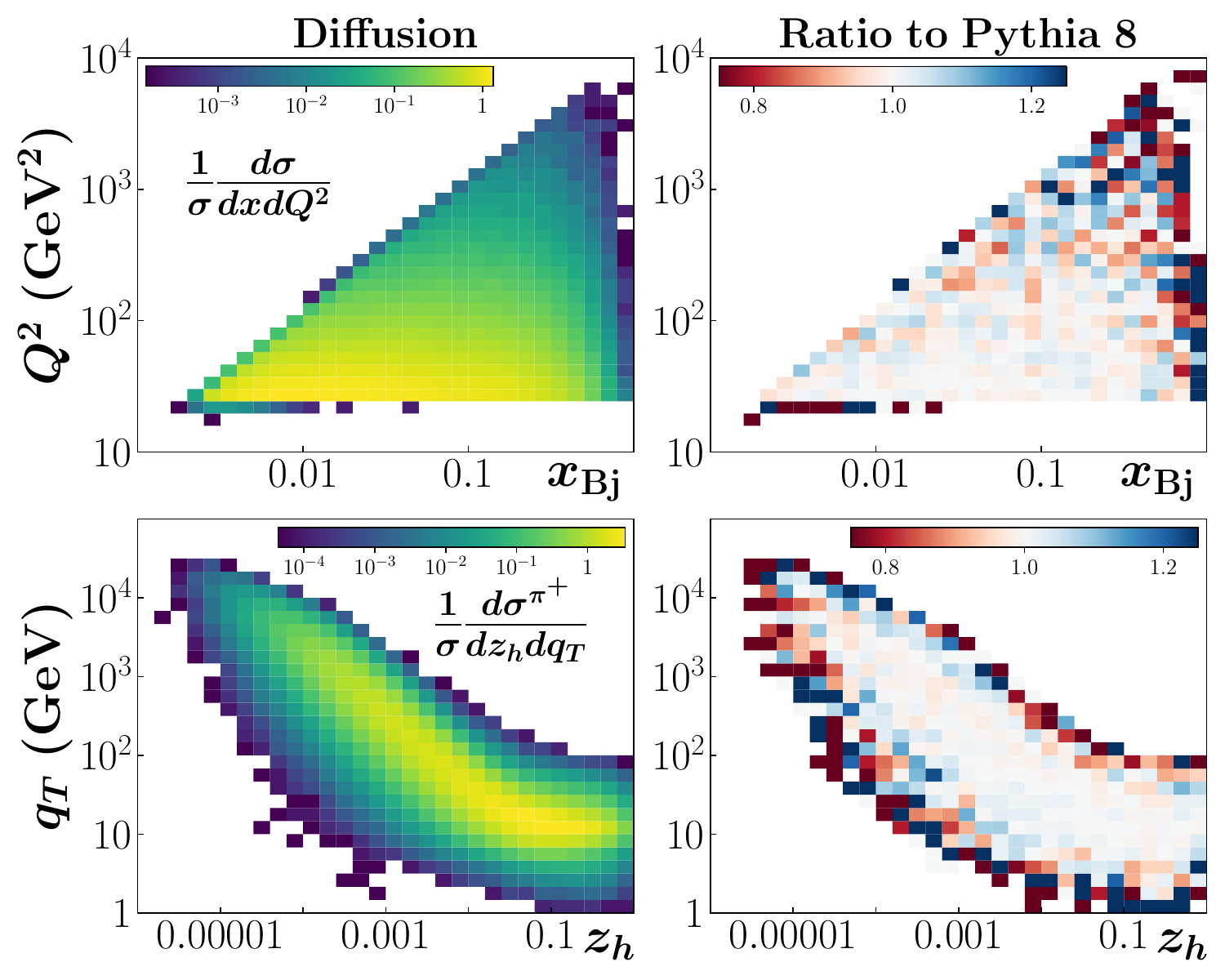}
    \caption{Top row: Diffusion model results and comparison to \textsc{Pythia8} for the DIS variables Bjorken $x$ and photon virtuality $Q^2$. Bottom row: Analogous comparison for the energy $z_{h}$ of pions and transverse momentum $q_T$ in the Breit frame relevant for SIDIS.}
    \label{fig:DIS}
\end{figure}

In the upper left panel of Fig.~\ref{fig:DIS}, we show the results from the diffusion model for the DIS variables in Eq.~(\ref{eq:DISVar}) as a two-dimensional histogram. In the upper right panel, we show a comparison of the diffusion model results for the DIS variables relative to \textsc{Pythia8}. We observe good agreement over the entire kinematic range. Minor deviations are noticeable only near the kinematic endpoints. Similar to the results for the kinematic distributions for the leading electron in Fig.~\ref{fig:DIS}, the deviations near the endpoint are likely due to statistical effects. In the lower two panels of Fig.~\ref{fig:DIS}, we show the analogous results for the SIDIS variables given in Eq.~(\ref{eq:SIDISVar}) for pions. Again, we find good agreement indicating the suitability of our model for different applications at the future EIC.

\subsection{Learning event-wide constraints}

\begin{table*}[!t]
    \setlength\tabcolsep{5pt}
    \centering
    \begin{tabular}{|c|c|c|c||c|c|c|}
    \cline{2-7}
          \multicolumn{1}{c|}{}          & \multicolumn{3}{c||}{Image-based diffusion model, Ref.~\cite{Devlin:2023jzp}} & \multicolumn{3}{c|}{Point cloud-based diffusion model (this work)} \\\cline{2-7}
          \multicolumn{1}{c|}{}  & $e^-$                         & $K^+$                 & $\pi^+$ &  $e^-$ & $K^+$ & $\pi^+$               \\\hline\hline
         $W_1^P(\eta)$ & $63.167 \pm 0.035$            & $36.669 \pm 0.029$  & $57.887 \pm 0.062$               & $0.266 \pm 0.009$   &$0.041 \pm 0.003$ & $0.310 \pm 0.016$\\
         $W_1^P(\phi)$ & $18.910 \pm 0.054$            & $18.736 \pm 0.048$  & $18.789 \pm 0.030$               & $0.015 \pm 0.004$   &$0.025 \pm 0.009$ &$0.158 \pm 0.003$\\
         $W_1^P(p_T)$  & $5.917 \pm 0.005$             & $0.323 \pm 0.002$   & $0.820 \pm 0.007$                & $0.251 \pm 0.004$   &$0.129 \pm 0.005$ &$0.464 \pm 0.007$\\
         Cov           & 0.011                         & 0.017               & 0.010                            & 0.546               &0.518             &0.473\\
         MMD           & 1.266                         & 2.160               & 1.945                            & 0.166               &0.382             &0.595\\
         KPD           & $7\times10^7 \pm 1\times10^7$ & $20.576 \pm 26.608$ & $4.6\times10^3 \pm 1.5\times10^3$& $0.0023 \pm 0.0003$ & 0                &$0.0062 \pm 0.0009$\\\hline
    \end{tabular}
    \caption{Metrics quantifying the performance of the image- and point cloud-based diffusion models compared to \textsc{Pythia8}. Small values are preferred for each metric except for the coverage.}
    \label{tab:metric}
\end{table*}

Generative models of full collider events need to satisfy global constraints such as momentum conservation; see Eq.~(\ref{eq:mtmconservation}) above. In addition, discrete quantum numbers such as the total baryon and lepton numbers need to be conserved. For each electron-proton event, we expect to have $\sum_{i\in{\rm event}} L_i = 1$ and $\sum_{i\in{\rm event}} B_i = 1$, where $L_i$ and $B_i$ are the lepton and baryon numbers of the $i^{\rm th}$ particle in the considered event, respectively. For electrons, muons, and neutrinos, we assign $L = +1$ and $L = -1$ for their antiparticles. All other particles are assigned $L = 0$. Analogous considerations apply to the baryon number. Due to experimental considerations, we combine neutrinos and anti-neutrinos as well as neutrons and anti-neutrons and exclude them when evaluating the event-wide lepton and baryon number. To assess the agreement between the diffusion model and \textsc{Pythia8}, we consider the ratio between the two for the momentum sum rule as well as the discrete quantum numbers within a 1-$\sigma$ confidence level:
\begin{align}
    \text{Momentum:}      \; 0.999(3)  \,,\nonumber\\
    \text{Baryon number:} \; 0.995(2) \,, \nonumber\\
    \text{Lepton number:} \; 1.001(2) \,. \nonumber
\end{align}
Overall, we observe very good agreement with deviations limited to the subpercent level.

The extent to which these conservation laws need to be satisfied depends on the specific application of the diffusion model. Alternatively, additional constraints could be incorporated into the training process, where violations are penalized, or the conservation laws can be strictly enforced on an event-by-event basis. We leave a quantitative comparison of different approaches for future work.

\subsection{Image vs. point cloud-based diffusion models}

To better evaluate the improvements achieved in this work compared to the image-based diffusion model from Ref.~\cite{Devlin:2023jzp} for electron-proton scattering events, we present the values for several quantitative metrics in Table~\ref{tab:metric}. We evaluate the models using the Wasserstein distance for the particle transverse momentum, rapidity, and azimuthal angle along with coverage (Cov), Maximum Mean Discrepancy (MMD) using the energy mover’s distance, and kernel physics distance (KPD). See Ref.~\cite{Kansal:2022spb} for more details. We only focus on the comparison for electrons, kaons, and pions, as the image-based diffusion model in Ref.~\cite{Devlin:2023jzp} was limited to these three particle species. Lower values of the different metrics indicate better results except for the coverage, where higher values are preferred. While some metrics show a more significant improvement than others, the point cloud-based model presented here consistently outperforms the image-based diffusion model of Ref.~\cite{Devlin:2023jzp}. This can be attributed to both its more advanced architecture and the loss of granularity in the image-based model due to pixelation.

\section{Conclusions and Outlook~\label{sec:conclusions}}

In this work, we introduced a diffusion model to generate full events at the future Electron-Ion Collider. Expanding on previous work, we developed a point cloud-based model combining edge creation with transformer modules to generate all particle species in the event. We evaluated the model's performance using different metrics and kinematic distributions, observing significant improvements across all metrics compared to earlier results. The model approximately learned event-wide momentum conservation, as well as the conservation of discrete quantum numbers such as baryon and lepton numbers. We expect that a similar multi-step generative process employed here could be applied to generate full events in other collision systems. By adopting the foundation model OmniLearn, our work may indicate a transition toward adapting foundation models for downstream tasks in fundamental particle and nuclear physics. In future work, we will explore different applications of the diffusion model developed here in the context of collider phenomenology, including fast simulations, inference tasks, and anomaly detection.

\section*{Code availability}

The code for this paper can be found at \url{https://github.com/ViniciusMikuni/OmniLearn} while the data generated to train the model is available at \url{https://zenodo.org/records/14027110}.

\begin{acknowledgements}
We would like to thank Kaori Fuyuto, Chris Lee, Emanuele Mereghetti and Benjamin Nachman for helpful discussions. JYA, FR, and NS were supported by the U.S. Department of Energy, Office of Science, Contract No.~DE-AC05-06OR23177, under which Jefferson Science Associates, LLC operates Jefferson Lab. JYA and FR were supported in part by the DOE, Office of Science, Office of Nuclear Physics, Early Career Program under contract No~DE-SC0024358. NS and RW are supported by the DOE, Office of Science, Office of Nuclear Physics in the Early Career Program. VM and FT are supported by the U.S. Department of Energy (DOE), Office of Science under contract DE-AC02-05CH11231. This research used resources of the National Energy Research Scientific Computing Center, a DOE Office of Science User Facility using NERSC award NERSC DDR-ERCAP0030239. This research was supported in part by the Quark-Gluon Tomography (QGT) Topical Collaboration, under contract no. DE-SC0023646.
\end{acknowledgements}

\bibliographystyle{utphys}
\bibliography{main.bib, other.bib}

\providecommand{\href}[2]{#2}\begingroup\raggedright\begin{thebibliography}{10}

\bibitem{AbdulKhalek:2021gbh}
R.~Abdul~Khalek {\em et~al.}, ``{Science Requirements and Detector Concepts for the Electron-Ion Collider: EIC Yellow Report},'' \href{http://arxiv.org/abs/2103.05419}{{\ttfamily arXiv:2103.05419 [physics.ins-det]}}.

\bibitem{2013arXiv1312.6114K}
D.~P. {Kingma} and M.~{Welling}, ``{Auto-Encoding Variational Bayes},'' \href{http://dx.doi.org/10.48550/arXiv.1312.6114}{{\em arXiv e-prints} (Dec., 2013) arXiv:1312.6114}, \href{http://arxiv.org/abs/1312.6114}{{\ttfamily arXiv:1312.6114 [stat.ML]}}.

\bibitem{Goodfellow:2014}
I.~J. Goodfellow, J.~Pouget-Abadie, M.~Mirza, B.~Xu, D.~Warde-Farley, S.~Ozair, A.~Courville, and Y.~Bengio, ``Generative adversarial nets,'' in {\em Proceedings of NIPS'14}, pp.~2672--2680.
\newblock Cambridge, MA, USA, 2014.
\newblock \url{http://dl.acm.org/citation.cfm?id=2969033.2969125}.

\bibitem{DBLP:journals/corr/Sohl-DicksteinW15}
J.~Sohl{-}Dickstein, E.~A. Weiss, N.~Maheswaranathan, and S.~Ganguli, ``Deep unsupervised learning using nonequilibrium thermodynamics,'' {\em CoRR} {\bfseries abs/1503.03585} (2015) , \href{http://arxiv.org/abs/1503.03585}{{\ttfamily 1503.03585}}. \url{http://arxiv.org/abs/1503.03585}.

\bibitem{DBLP:journals/corr/abs-2006-11239}
J.~Ho, A.~Jain, and P.~Abbeel, ``Denoising diffusion probabilistic models,'' {\em CoRR} {\bfseries abs/2006.11239} (2020) , \href{http://arxiv.org/abs/2006.11239}{{\ttfamily 2006.11239}}. \url{https://arxiv.org/abs/2006.11239}.

\bibitem{Kasieczka:2017nvn}
G.~Kasieczka, T.~Plehn, M.~Russell, and T.~Schell, ``{Deep-learning Top Taggers or The End of QCD?},'' \href{http://dx.doi.org/10.1007/JHEP05(2017)006}{{\em JHEP} {\bfseries 05} (2017) 006}, \href{http://arxiv.org/abs/1701.08784}{{\ttfamily arXiv:1701.08784 [hep-ph]}}.

\bibitem{Cai:2021hnn}
T.~Cai, J.~Cheng, K.~Craig, and N.~Craig, ``{Which metric on the space of collider events?},'' \href{http://dx.doi.org/10.1103/PhysRevD.105.076003}{{\em Phys. Rev. D} {\bfseries 105} no.~7, (2022) 076003}, \href{http://arxiv.org/abs/2111.03670}{{\ttfamily arXiv:2111.03670 [hep-ph]}}.

\bibitem{Datta:2017rhs}
K.~Datta and A.~Larkoski, ``{How Much Information is in a Jet?},'' \href{http://dx.doi.org/10.1007/JHEP06(2017)073}{{\em JHEP} {\bfseries 06} (2017) 073}, \href{http://arxiv.org/abs/1704.08249}{{\ttfamily arXiv:1704.08249 [hep-ph]}}.

\bibitem{Komiske:2018cqr}
P.~T. Komiske, E.~M. Metodiev, and J.~Thaler, ``{Energy Flow Networks: Deep Sets for Particle Jets},'' \href{http://dx.doi.org/10.1007/JHEP01(2019)121}{{\em JHEP} {\bfseries 01} (2019) 121}, \href{http://arxiv.org/abs/1810.05165}{{\ttfamily arXiv:1810.05165 [hep-ph]}}.

\bibitem{Heimel:2018mkt}
T.~Heimel, G.~Kasieczka, T.~Plehn, and J.~M. Thompson, ``{QCD or What?},'' \href{http://dx.doi.org/10.21468/SciPostPhys.6.3.030}{{\em SciPost Phys.} {\bfseries 6} no.~3, (2019) 030}, \href{http://arxiv.org/abs/1808.08979}{{\ttfamily arXiv:1808.08979 [hep-ph]}}.

\bibitem{Dreyer:2021hhr}
F.~A. Dreyer, G.~Soyez, and A.~Takacs, ``{Quarks and gluons in the Lund plane},'' \href{http://dx.doi.org/10.1007/JHEP08(2022)177}{{\em JHEP} {\bfseries 08} (2022) 177}, \href{http://arxiv.org/abs/2112.09140}{{\ttfamily arXiv:2112.09140 [hep-ph]}}.

\bibitem{Bellagente:2019uyp}
M.~Bellagente, A.~Butter, G.~Kasieczka, T.~Plehn, and R.~Winterhalder, ``{How to GAN away Detector Effects},'' \href{http://dx.doi.org/10.21468/SciPostPhys.8.4.070}{{\em SciPost Phys.} {\bfseries 8} no.~4, (2020) 070}, \href{http://arxiv.org/abs/1912.00477}{{\ttfamily arXiv:1912.00477 [hep-ph]}}.

\bibitem{Andreassen:2019cjw}
A.~Andreassen, P.~T. Komiske, E.~M. Metodiev, B.~Nachman, and J.~Thaler, ``{OmniFold: A Method to Simultaneously Unfold All Observables},'' \href{http://dx.doi.org/10.1103/PhysRevLett.124.182001}{{\em Phys. Rev. Lett.} {\bfseries 124} no.~18, (2020) 182001}, \href{http://arxiv.org/abs/1911.09107}{{\ttfamily arXiv:1911.09107 [hep-ph]}}.

\bibitem{Alanazi:2020jod}
Y.~Alanazi {\em et~al.}, ``{Machine learning-based event generator for electron-proton scattering},'' \href{http://dx.doi.org/10.1103/PhysRevD.106.096002}{{\em Phys. Rev. D} {\bfseries 106} no.~9, (2022) 096002}, \href{http://arxiv.org/abs/2008.03151}{{\ttfamily arXiv:2008.03151 [hep-ph]}}.

\bibitem{Alghamdi:2023emm}
T.~Alghamdi {\em et~al.}, ``{Toward a generative modeling analysis of CLAS exclusive 2\ensuremath{\pi} photoproduction},'' \href{http://dx.doi.org/10.1103/PhysRevD.108.094030}{{\em Phys. Rev. D} {\bfseries 108} no.~9, (2023) 094030}, \href{http://arxiv.org/abs/2307.04450}{{\ttfamily arXiv:2307.04450 [hep-ph]}}.

\bibitem{Huang:2023kgs}
Y.~Huang, D.~Torbunov, B.~Viren, H.~Yu, J.~Huang, M.~Lin, and Y.~Ren, ``{Unsupervised Domain Transfer for Science: Exploring Deep Learning Methods for Translation between LArTPC Detector Simulations with Differing Response Models},'' \href{http://arxiv.org/abs/2304.12858}{{\ttfamily arXiv:2304.12858 [hep-ex]}}.

\bibitem{Lee:2022kdn}
K.~Lee, J.~Mulligan, M.~P\l{}osko\'n, F.~Ringer, and F.~Yuan, ``{Machine learning-based jet and event classification at the Electron-Ion Collider with applications to hadron structure and spin physics},'' \href{http://dx.doi.org/10.1007/JHEP03(2023)085}{{\em JHEP} {\bfseries 03} (2023) 085}, \href{http://arxiv.org/abs/2210.06450}{{\ttfamily arXiv:2210.06450 [hep-ph]}}.

\bibitem{Butter:2019cae}
A.~Butter, T.~Plehn, and R.~Winterhalder, ``{How to GAN LHC Events},'' \href{http://dx.doi.org/10.21468/SciPostPhys.7.6.075}{{\em SciPost Phys.} {\bfseries 7} no.~6, (2019) 075}, \href{http://arxiv.org/abs/1907.03764}{{\ttfamily arXiv:1907.03764 [hep-ph]}}.

\bibitem{Gao:2020zvv}
C.~Gao, S.~H\"oche, J.~Isaacson, C.~Krause, and H.~Schulz, ``{Event Generation with Normalizing Flows},'' \href{http://dx.doi.org/10.1103/PhysRevD.101.076002}{{\em Phys. Rev. D} {\bfseries 101} no.~7, (2020) 076002}, \href{http://arxiv.org/abs/2001.10028}{{\ttfamily arXiv:2001.10028 [hep-ph]}}.

\bibitem{Danziger:2021eeg}
K.~Danziger, T.~Jan\ss{}en, S.~Schumann, and F.~Siegert, ``{Accelerating Monte Carlo event generation -- rejection sampling using neural network event-weight estimates},'' \href{http://dx.doi.org/10.21468/SciPostPhys.12.5.164}{{\em SciPost Phys.} {\bfseries 12} (2022) 164}, \href{http://arxiv.org/abs/2109.11964}{{\ttfamily arXiv:2109.11964 [hep-ph]}}.

\bibitem{Butter:2022rso}
S.~Badger {\em et~al.}, ``{Machine learning and LHC event generation},'' \href{http://dx.doi.org/10.21468/SciPostPhys.14.4.079}{{\em SciPost Phys.} {\bfseries 14} no.~4, (2023) 079}, \href{http://arxiv.org/abs/2203.07460}{{\ttfamily arXiv:2203.07460 [hep-ph]}}.

\bibitem{Cirigliano:2021img}
V.~Cirigliano, K.~Fuyuto, C.~Lee, E.~Mereghetti, and B.~Yan, ``{Charged Lepton Flavor Violation at the EIC},'' \href{http://dx.doi.org/10.1007/JHEP03(2021)256}{{\em JHEP} {\bfseries 03} (2021) 256}, \href{http://arxiv.org/abs/2102.06176}{{\ttfamily arXiv:2102.06176 [hep-ph]}}.

\bibitem{Nachman:2020lpy}
B.~Nachman and D.~Shih, ``{Anomaly Detection with Density Estimation},'' \href{http://dx.doi.org/10.1103/PhysRevD.101.075042}{{\em Phys. Rev. D} {\bfseries 101} (2020) 075042}, \href{http://arxiv.org/abs/2001.04990}{{\ttfamily arXiv:2001.04990 [hep-ph]}}.

\bibitem{Atkinson:2022uzb}
O.~Atkinson, A.~Bhardwaj, C.~Englert, P.~Konar, V.~S. Ngairangbam, and M.~Spannowsky, ``{IRC-Safe Graph Autoencoder for Unsupervised Anomaly Detection},'' \href{http://dx.doi.org/10.3389/frai.2022.943135}{{\em Front. Artif. Intell.} {\bfseries 5} (2022) 943135}, \href{http://arxiv.org/abs/2204.12231}{{\ttfamily arXiv:2204.12231 [hep-ph]}}.

\bibitem{Scheinker:2024anx}
A.~Scheinker, ``{cDVAE: Multimodal Generative Conditional Diffusion Guided by Variational Autoencoder Latent Embedding for Virtual 6D Phase Space Diagnostics},'' \href{http://arxiv.org/abs/2407.20218}{{\ttfamily arXiv:2407.20218 [physics.acc-ph]}}.

\bibitem{Andreassen:2020nkr}
A.~Andreassen, B.~Nachman, and D.~Shih, ``{Simulation Assisted Likelihood-free Anomaly Detection},'' \href{http://dx.doi.org/10.1103/PhysRevD.101.095004}{{\em Phys. Rev. D} {\bfseries 101} no.~9, (2020) 095004}, \href{http://arxiv.org/abs/2001.05001}{{\ttfamily arXiv:2001.05001 [hep-ph]}}.

\bibitem{Finke:2021sdf}
T.~Finke, M.~Kr\"amer, A.~Morandini, A.~M\"uck, and I.~Oleksiyuk, ``{Autoencoders for unsupervised anomaly detection in high energy physics},'' \href{http://dx.doi.org/10.1007/JHEP06(2021)161}{{\em JHEP} {\bfseries 06} (2021) 161}, \href{http://arxiv.org/abs/2104.09051}{{\ttfamily arXiv:2104.09051 [hep-ph]}}.

\bibitem{Fraser:2021lxm}
K.~Fraser, S.~Homiller, R.~K. Mishra, B.~Ostdiek, and M.~D. Schwartz, ``{Challenges for unsupervised anomaly detection in particle physics},'' \href{http://dx.doi.org/10.1007/JHEP03(2022)066}{{\em JHEP} {\bfseries 03} (2022) 066}, \href{http://arxiv.org/abs/2110.06948}{{\ttfamily arXiv:2110.06948 [hep-ph]}}.

\bibitem{Araz:2022zxk}
J.~Y. Araz and M.~Spannowsky, ``{Quantum-probabilistic Hamiltonian learning for generative modelling \& anomaly detection},'' \href{http://arxiv.org/abs/2211.03803}{{\ttfamily arXiv:2211.03803 [quant-ph]}}.

\bibitem{Sengupta:2023vtm}
D.~Sengupta, M.~Leigh, J.~A. Raine, S.~Klein, and T.~Golling, ``{Improving new physics searches with diffusion models for event observables and jet constituents},'' \href{http://dx.doi.org/10.1007/JHEP04(2024)109}{{\em JHEP} {\bfseries 04} (2024) 109}, \href{http://arxiv.org/abs/2312.10130}{{\ttfamily arXiv:2312.10130 [physics.data-an]}}.

\bibitem{Morandini:2023pwj}
A.~Morandini, T.~Ferber, and F.~Kahlhoefer, ``{Reconstructing axion-like particles from beam dumps with simulation-based inference},'' \href{http://arxiv.org/abs/2308.01353}{{\ttfamily arXiv:2308.01353 [hep-ph]}}.

\bibitem{Birk:2024knn}
J.~Birk, A.~Hallin, and G.~Kasieczka, ``{OmniJet-\ensuremath{\alpha}: the first cross-task foundation model for particle physics},'' \href{http://dx.doi.org/10.1088/2632-2153/ad66ad}{{\em Mach. Learn. Sci. Tech.} {\bfseries 5} no.~3, (2024) 035031}, \href{http://arxiv.org/abs/2403.05618}{{\ttfamily arXiv:2403.05618 [hep-ph]}}.

\bibitem{Mikuni:2024qsr}
V.~Mikuni and B.~Nachman, ``{OmniLearn: A Method to Simultaneously Facilitate All Jet Physics Tasks},'' \href{http://arxiv.org/abs/2404.16091}{{\ttfamily arXiv:2404.16091 [hep-ph]}}.

\bibitem{deOliveira:2017pjk}
L.~de~Oliveira, M.~Paganini, and B.~Nachman, ``{Learning Particle Physics by Example: Location-Aware Generative Adversarial Networks for Physics Synthesis},'' \href{http://dx.doi.org/10.1007/s41781-017-0004-6}{{\em Comput. Softw. Big Sci.} {\bfseries 1} no.~1, (2017) 4}, \href{http://arxiv.org/abs/1701.05927}{{\ttfamily arXiv:1701.05927 [stat.ML]}}.

\bibitem{Paganini:2017dwg}
M.~Paganini, L.~de~Oliveira, and B.~Nachman, ``{CaloGAN : Simulating 3D high energy particle showers in multilayer electromagnetic calorimeters with generative adversarial networks},'' \href{http://dx.doi.org/10.1103/PhysRevD.97.014021}{{\em Phys. Rev. D} {\bfseries 97} no.~1, (2018) 014021}, \href{http://arxiv.org/abs/1712.10321}{{\ttfamily arXiv:1712.10321 [hep-ex]}}.

\bibitem{Alanazi:2020klf}
Y.~Alanazi {\em et~al.}, ``{Simulation of electron-proton scattering events by a Feature-Augmented and Transformed Generative Adversarial Network (FAT-GAN)},'' \href{http://arxiv.org/abs/2001.11103}{{\ttfamily arXiv:2001.11103 [hep-ph]}}.

\bibitem{Kansal:2021cqp}
R.~Kansal, J.~Duarte, H.~Su, B.~Orzari, T.~Tomei, M.~Pierini, M.~Touranakou, J.-R. Vlimant, and D.~Gunopulos, ``{Particle Cloud Generation with Message Passing Generative Adversarial Networks},'' \href{http://arxiv.org/abs/2106.11535}{{\ttfamily arXiv:2106.11535 [cs.LG]}}.

\bibitem{Buhmann:2023pmh}
E.~Buhmann, G.~Kasieczka, and J.~Thaler, ``{EPiC-GAN: Equivariant Point Cloud Generation for Particle Jets},'' \href{http://arxiv.org/abs/2301.08128}{{\ttfamily arXiv:2301.08128 [hep-ph]}}.

\bibitem{Touranakou:2022qrp}
M.~Touranakou, N.~Chernyavskaya, J.~Duarte, D.~Gunopulos, R.~Kansal, B.~Orzari, M.~Pierini, T.~Tomei, and J.-R. Vlimant, ``{Particle-based Fast Jet Simulation at the LHC with Variational Autoencoders},'' \href{http://dx.doi.org/10.1088/2632-2153/ac7c56}{{\em Mach.Learn.Sci.Tech.} {\bfseries 3} (3, 2022) 035003}, \href{http://arxiv.org/abs/2203.00520}{{\ttfamily arXiv:2203.00520 [physics.comp-ph]}}.

\bibitem{Kach:2022qnf}
B.~K\"ach, D.~Kr\"ucker, I.~Melzer-Pellmann, M.~Scham, S.~Schnake, and A.~Verney-Provatas, ``{JetFlow: Generating Jets with Conditioned and Mass Constrained Normalising Flows},'' \href{http://arxiv.org/abs/2211.13630}{{\ttfamily arXiv:2211.13630 [hep-ex]}}.

\bibitem{Verheyen:2022tov}
R.~Verheyen, ``{Event Generation and Density Estimation with Surjective Normalizing Flows},'' \href{http://dx.doi.org/10.21468/SciPostPhys.13.3.047}{{\em SciPost Phys.} {\bfseries 13} (5, 2022) 047}, \href{http://arxiv.org/abs/2205.01697}{{\ttfamily arXiv:2205.01697 [hep-ph]}}.

\bibitem{DBLP:journals/corr/abs-1907-05600}
Y.~Song and S.~Ermon, ``Generative modeling by estimating gradients of the data distribution,'' {\em CoRR} {\bfseries abs/1907.05600} (2019) , \href{http://arxiv.org/abs/1907.05600}{{\ttfamily 1907.05600}}. \url{http://arxiv.org/abs/1907.05600}.

\bibitem{Mikuni:2022xry}
V.~Mikuni and B.~Nachman, ``{Score-based generative models for calorimeter shower simulation},'' \href{http://dx.doi.org/10.1103/PhysRevD.106.092009}{{\em Phys. Rev. D} {\bfseries 106} no.~9, (2022) 092009}, \href{http://arxiv.org/abs/2206.11898}{{\ttfamily arXiv:2206.11898 [hep-ph]}}.

\bibitem{Mikuni:2023dvk}
V.~Mikuni, B.~Nachman, and M.~Pettee, ``{Fast Point Cloud Generation with Diffusion Models in High Energy Physics},'' \href{http://arxiv.org/abs/2304.01266}{{\ttfamily arXiv:2304.01266 [hep-ph]}}.

\bibitem{Leigh:2023toe}
M.~Leigh, D.~Sengupta, G.~Qu\'etant, J.~A. Raine, K.~Zoch, and T.~Golling, ``{PC-JeDi: Diffusion for Particle Cloud Generation in High Energy Physics},'' \href{http://arxiv.org/abs/2303.05376}{{\ttfamily arXiv:2303.05376 [hep-ph]}}.

\bibitem{Butter:2023fov}
A.~Butter, N.~Huetsch, S.~P. Schweitzer, T.~Plehn, P.~Sorrenson, and J.~Spinner, ``{Jet Diffusion versus JetGPT -- Modern Networks for the LHC},'' \href{http://arxiv.org/abs/2305.10475}{{\ttfamily arXiv:2305.10475 [hep-ph]}}.

\bibitem{Acosta:2023zik}
F.~T. Acosta, V.~Mikuni, B.~Nachman, M.~Arratia, K.~Barish, B.~Karki, R.~Milton, P.~Karande, and A.~Angerami, ``{Comparison of Point Cloud and Image-based Models for Calorimeter Fast Simulation},'' \href{http://arxiv.org/abs/2307.04780}{{\ttfamily arXiv:2307.04780 [cs.LG]}}.

\bibitem{Leigh:2023zle}
M.~Leigh, D.~Sengupta, J.~A. Raine, G.~Qu\'etant, and T.~Golling, ``{PC-Droid: Faster diffusion and improved quality for particle cloud generation},'' \href{http://arxiv.org/abs/2307.06836}{{\ttfamily arXiv:2307.06836 [hep-ex]}}.

\bibitem{Amram:2023onf}
O.~Amram and K.~Pedro, ``{Denoising diffusion models with geometry adaptation for high fidelity calorimeter simulation},'' \href{http://arxiv.org/abs/2308.03876}{{\ttfamily arXiv:2308.03876 [physics.ins-det]}}.

\bibitem{Buhmann:2023zgc}
E.~Buhmann, C.~Ewen, D.~A. Faroughy, T.~Golling, G.~Kasieczka, M.~Leigh, G.~Qu\'etant, J.~A. Raine, D.~Sengupta, and D.~Shih, ``{EPiC-ly Fast Particle Cloud Generation with Flow-Matching and Diffusion},'' \href{http://arxiv.org/abs/2310.00049}{{\ttfamily arXiv:2310.00049 [hep-ph]}}.

\bibitem{Imani:2023blb}
Z.~Imani, S.~Aeron, and T.~Wongjirad, ``{Score-based Diffusion Models for Generating Liquid Argon Time Projection Chamber Images},'' \href{http://arxiv.org/abs/2307.13687}{{\ttfamily arXiv:2307.13687 [hep-ex]}}.

\bibitem{Mikuni:2023tqg}
V.~Mikuni and B.~Nachman, ``{CaloScore v2: Single-shot Calorimeter Shower Simulation with Diffusion Models},'' \href{http://arxiv.org/abs/2308.03847}{{\ttfamily arXiv:2308.03847 [hep-ph]}}.

\bibitem{Devlin:2023jzp}
P.~Devlin, J.-W. Qiu, F.~Ringer, and N.~Sato, ``{Diffusion model approach to simulating electron-proton scattering events},'' \href{http://dx.doi.org/10.1103/PhysRevD.110.016030}{{\em Phys. Rev. D} {\bfseries 110} no.~1, (2024) 016030}, \href{http://arxiv.org/abs/2310.16308}{{\ttfamily arXiv:2310.16308 [hep-ph]}}.

\bibitem{Go:2024xor}
Y.~Go, D.~Torbunov, T.~Rinn, Y.~Huang, H.~Yu, B.~Viren, M.~Lin, Y.~Ren, and J.~Huang, ``{Effectiveness of denoising diffusion probabilistic models for fast and high-fidelity whole-event simulation in high-energy heavy-ion experiments},'' \href{http://arxiv.org/abs/2406.01602}{{\ttfamily arXiv:2406.01602 [physics.data-an]}}.

\bibitem{Sjostrand:2014zea}
T.~Sj{\"o}strand, S.~Ask, J.~R. Christiansen, R.~Corke, N.~Desai, P.~Ilten, S.~Mrenna, S.~Prestel, C.~O. Rasmussen, and P.~Z. Skands, ``{An Introduction to PYTHIA 8.2},'' \href{http://dx.doi.org/10.1016/j.cpc.2015.01.024}{{\em Comput. Phys. Commun.} {\bfseries 191} (2015) 159--177},
\href{http://arxiv.org/abs/1410.3012}{{\ttfamily arXiv:1410.3012 [hep-ph]}}.

\bibitem{tancik2020fourier}
M.~Tancik, P.~Srinivasan, B.~Mildenhall, S.~Fridovich-Keil, N.~Raghavan, U.~Singhal, R.~Ramamoorthi, J.~Barron, and R.~Ng, ``Fourier features let networks learn high frequency functions in low dimensional domains,'' in {\em Advances in Neural Information Processing Systems}, H.~Larochelle, M.~Ranzato, R.~Hadsell, M.~Balcan, and H.~Lin, eds., vol.~33, pp.~7537--7547.
\newblock Curran Associates, Inc., 2020.
\newblock \url{https://proceedings.neurips.cc/paper_files/paper/2020/file/55053683268957697aa39fba6f231c68-Paper.pdf}.

\bibitem{hendrycks2016gaussian}
D.~Hendrycks and K.~Gimpel, ``Gaussian error linear units (gelus),'' {\em arXiv preprint arXiv:1606.08415} (2016) .

\bibitem{he2016deep}
K.~He, X.~Zhang, S.~Ren, and J.~Sun, ``Deep residual learning for image recognition,'' in {\em Proceedings of the IEEE conference on computer vision and pattern recognition}, pp.~770--778.
\newblock 2016.

\bibitem{Mikuni:2021pou}
V.~Mikuni and F.~Canelli, ``{Point cloud transformers applied to collider physics},'' \href{http://dx.doi.org/10.1088/2632-2153/ac07f6}{{\em Mach. Learn. Sci. Tech.} {\bfseries 2} no.~3, (2021) 035027}, \href{http://arxiv.org/abs/2102.05073}{{\ttfamily arXiv:2102.05073 [physics.data-an]}}.

\bibitem{DBLP:journals/corr/abs-1801-07829}
Y.~Wang, Y.~Sun, Z.~Liu, S.~E. Sarma, M.~M. Bronstein, and J.~M. Solomon, ``Dynamic graph {CNN} for learning on point clouds,'' {\em CoRR} {\bfseries abs/1801.07829} (2018) , \href{http://arxiv.org/abs/1801.07829}{{\ttfamily 1801.07829}}. \url{http://arxiv.org/abs/1801.07829}.

\bibitem{Perlmutter}
``{Perlmutter} system.'' \url{https://docs.nersc.gov/systems/perlmutter/system_details/}.
\newblock Accessed: 2022-05-04.

\bibitem{sergeev2018horovod}
A.~Sergeev and M.~D. Balso, ``Horovod: fast and easy distributed deep learning in {TensorFlow},'' {\em arXiv preprint arXiv:1802.05799} (2018) .

\bibitem{tensorflow}
M.~Abadi, P.~Barham, J.~Chen, Z.~Chen, A.~Davis, J.~Dean, M.~Devin, S.~Ghemawat, G.~Irving, M.~Isard, {\em et~al.}, ``Tensorflow: A system for large-scale machine learning.,'' in {\em OSDI}, vol.~16, pp.~265--283.
\newblock 2016.

\bibitem{keras}
F.~Chollet, ``Keras.'' \url{https://github.com/fchollet/keras}, 2017.

\bibitem{DBLP:journals/corr/LoshchilovH16a}
I.~Loshchilov and F.~Hutter, ``{SGDR:} stochastic gradient descent with restarts,'' {\em CoRR} {\bfseries abs/1608.03983} (2016) , \href{http://arxiv.org/abs/1608.03983}{{\ttfamily 1608.03983}}. \url{http://arxiv.org/abs/1608.03983}.

\bibitem{chen2024symbolic}
X.~Chen, C.~Liang, D.~Huang, E.~Real, K.~Wang, H.~Pham, X.~Dong, T.~Luong, C.-J. Hsieh, Y.~Lu, {\em et~al.}, ``Symbolic discovery of optimization algorithms,'' {\em Advances in Neural Information Processing Systems} {\bfseries 36} (2024) .

\bibitem{Bacchetta:2006tn}
A.~Bacchetta, M.~Diehl, K.~Goeke, A.~Metz, P.~J. Mulders, and M.~Schlegel, ``{Semi-inclusive deep inelastic scattering at small transverse momentum},'' \href{http://dx.doi.org/10.1088/1126-6708/2007/02/093}{{\em JHEP} {\bfseries 02} (2007) 093}, \href{http://arxiv.org/abs/hep-ph/0611265}{{\ttfamily arXiv:hep-ph/0611265}}.

\bibitem{Kansal:2022spb}
R.~Kansal, A.~Li, J.~Duarte, N.~Chernyavskaya, M.~Pierini, B.~Orzari, and T.~Tomei, ``{Evaluating generative models in high energy physics},'' \href{http://dx.doi.org/10.1103/PhysRevD.107.076017}{{\em Phys. Rev. D} {\bfseries 107} no.~7, (2023) 076017}, \href{http://arxiv.org/abs/2211.10295}{{\ttfamily arXiv:2211.10295 [hep-ex]}}.

\end{thebibliography}\endgroup

\end{document}